\documentclass[a4paper,12pt,useAMS,usenatbib,revtex4]{mn2e}
\usepackage{lineno}
\usepackage{graphicx}
\usepackage{longtable}
\usepackage{amssymb}
\usepackage{tabularx, blindtext}
\usepackage{lscape}
\usepackage{url}
\title[Choosing the best law for optimal retrieval of transit parameters]
{Limb-darkening and exoplanets II: Choosing the Best Law for Optimal Retrieval of Transit Parameters}
\author[Espinoza \& Jord\'an ]{
N\'estor Espinoza$^{1,2}$\thanks{E-mail:nespino@astro.puc.cl}, 
Andr\'es Jord\'an$^{1,2}$\thanks{E-mail:ajordan@astro.puc.cl}\\ 
$^{1}$Instituto de Astrof\'isica, Pontificia Universidad Cat\'olica de Chile, Vicu\~na Mackenna 
4860, Santiago, Chile\\
$^{2}$Millennium Institute of Astrophysics, Santiago, Chile\\
}
\begin{document}

\date{}

\pagerange{\pageref{firstpage}--\pageref{lastpage}} \pubyear{2015}

\maketitle

\label{firstpage}

\begin{abstract}
Very precise measurements of exoplanet transit lightcurves both from ground and space based observatories make it now possible to fit the limb-darkening coefficients in the transit-fitting procedure rather than fix them to theoretical values. 
This strategy has been shown to give better results, as fixing the coefficients to theoretical values can give rise to important systematic 
errors which directly impact the physical properties of the system derived from such lightcurves such as the planetary radius. However, studies of the effect of limb darkening assumptions on the retrieved parameters 
have mostly focused on the widely used quadratic limb-darkening law, leaving out other proposed laws that are either simpler or 
better descriptions of model intensity profiles. In this work, we show that laws such as the 
logarithmic, square-root and three-parameter law do a better job than the quadratic and linear laws when deriving parameters from transit lightcurves, both in terms of 
bias and precision, for a wide range of situations. We therefore recommend to study which law to 
use on a case-by-case basis. We provide code to guide the decision of when to use each of these laws and select the optimal one in a 
mean-square error sense, which we note has a dependence on both stellar and transit parameters. Finally, we demonstrate that the 
so-called exponential law is non-physical as it typically produces negative intensities close to the limb and should therefore not be used.
\end{abstract}


% #################################################################################################################################
% #################################################################################################################################
% #################################################################################################################################
% ############################################# SECTION 1 #########################################################################
% #################################################################################################################################
% #################################################################################################################################
% #################################################################################################################################

\section{Introduction}

In the past decade, the study of transiting exoplanets has evolved from discovery to very precise characterization of these systems, 
thanks to the exquisite precision allowed mainly by space-based observatories such as the {\em Hubble Space Telescope} ({\em HST}) and the {\em Kepler} mission \citep{koch2010}.
This has allowed the study of precise mass-radius diagrams \citep[see, e.g.,][]{wm2014,dressing2015,WL2015b}, and in turn  allow the possibility of  obtaining a precise determination of the internal composition of small, rocky exoplanets based on mass and radius
measurements \citep{dorn2015} and of their atmospheric fractions from radius measurements alone \citep{WL2015}. In addition, precise transit lightcurves enable the determination of
derived parameters of the systems that even allow for the estimation of stellar parameters directly from transit lightcurves through techniques such as asterodensity
profiling \citep{sm2003,kipping2014AP}, and the detection of atmospheric features in exoplanet atmospheres through the technique of transmission spectroscopy. 
All these studies and techniques rely on the retrieval of {\em precise and accurate} transit parameters from transit lightcurves.

In a recent study \citep[][hereon EJ15]{ej2015}, we showed that the accuracy is actually catching up with the precision 
of these measurements due to our imperfect understanding of the limb-darkening effect. In particular, we showed that there are important 
biases both when fixing and fitting for the limb-darkening coefficients in the light curve fitting process. The biases when fitting for the 
limb darkening coefficients arise from the fact that the popular and widely used quadratic law is unable to model the complex intensity profile 
of real stars \citep[which has also been seen on other studies, see, e.g.,][]{knutson2007, kreidberg2015},  and the biases when fixing the coefficients arise for the
same reason plus the facts that (1) different methods of fitting the model intensity profiles (including transit light curve versus intensity profile fitting) give rise to 
different limb-darkening coefficients and (2) that we have an imperfect knowledge of the real intensity profiles of stars. We showed that fixing the limb-darkening 
coefficients is the worst option if one aims to obtain precise \textit{and} accurate transit parameters, because the biases arise from three different sources, 
with the last one (the fact that we do not have a perfect modelling of the stellar intensity profiles of real stars) having an unknown but possibly 
large impact.

Given the above, unless the data quality is really poor (and one is willing to trade bias for variance), there is actually no good reason to fix the limb-darkening coefficients 
in the transit fitting process (assuming computational resources are not an issue). However, although fitting the 
limb-darkening coefficients seems to be a good solution to the accuracy problem, there remains the issue of the low flexibility of the quadratic limb-darkening 
law which can cause important biases on the retrieved transit parameters as shown in EJ15. These biases can be as large as $\sim 1\%$ for the planet-to-star 
radius ratio $R_p/R_*$, $\sim 2\%$ for the scaled semi-major axis, $a/R_*$, and $2\%$ for the inclination, where $R_p$ is the planetary radius, $R_*$ 
the stellar radius and $a$ the semi-major axis. The importance for missions like {\em Kepler} and for future missions like the Transiting Exoplanet Survey 
Satellite \cite[TESS,][]{ricker2014} is evident when looking at the most recent results from the {\em Kepler} mission: if we focus on the planet-to-star radius ratio measurements 
alone, which according to EJ15 have an accuracy bias on the order of $\sim 0.2\%$, a query to the Nasa Exoplanet Archive\footnote{\url{http://exoplanetarchive.ipac.caltech.edu/}; query done on 29/09/2015.} shows that out of 3063 planetary candidates, 933 ($26\%$) have precisions better than 
$0.2\%$, and out of 1001 {\em Kepler} confirmed exoplanets, 463 (46\%) do too. This means that at least\footnote{It is important to note that this is an 
underestimate of the number of planet candidates with estimation errors, due to the fact that the Kepler pipeline makes use of fixed limb-darkening 
coefficients using the quadratic law \citep{rowe2015}, which as explained here and estimated in EJ15, gives rise to more severe biases than the one used for this 
estimation} $26\%$ of the planet candidates (from which, e.g., population studies, which rely on averaging out \textit{random} and not \textit{systematic} 
uncertainties like the ones introduced by limb-darkening) and almost half of the confirmed exoplanets have significant systematic errors.

An obvious solution to the issues mentioned above if one decides to fit the limb-darkening coefficients is to try to use alternative laws to describe 
the intensity profile of stars. Although the non-linear law proposed by \cite{claret2000} seems to be the most flexible, the fact that it has four free 
parameters does not make it a very attractive one given its high complexity. Several laws with fewer parameters 
have been proposed. Two-parameter laws are the exponential proposed by \cite{claret2003}, the logarithmic proposed by \cite{klinglesmith1970} and 
the square-root proposed by \cite{diazgimenez1992}. In addition, a very flexible three-parameter law was also proposed by \cite{sing2009}. These laws 
with fewer parameters are potentially very attractive for transit fitting purposes, due to their flexibility at following different center to limb stellar intensity profiles and their low 
number of parameters \citep[see, e.g., ][for a comparison between the goodness of fit to ATLAS model atmospheres of the mentioned two-parameter laws 
which outperform the linear and quadratic laws in terms of following the intensity profiles]{howarth2011}.

Despite the attractive nature of the above mentioned limb-darkening laws, testing their performance at retrieving transit parameters from transit 
lightcurves was problematic until recently for two reasons. First, there was no publicly available code capable of efficiently generating fast and accurate 
transit lightcurves using all of these non-standard laws. However, recently \cite{batman2015} published an 
algorithm that does exactly this called \texttt{batman}, enabling the generation of transit lightcurves very efficiently with any arbitrary limb-darkening law. 
The second problem was that it was not clear how to sample limb-darkening coefficients in an informative (i.e., sampling all the physically possible parameter 
space) and efficient way for all of these laws. Recently, \cite{kipping2015} presented an algorithm to sample parameters from the three-parameter law by imposing 
physically motivated constraints on the intensity profiles which derived in constrains on the limb-darkening coefficients being fitted. \cite{kipping2013}, using the 
same principles, derived an algorithmic way of doing this for two-parameter laws. He found that his algorithm was not applicable to the exponential law and thus 
no constrains for this law could be derived. We note that the form of the logarithmic law used in \cite{kipping2013} differs from the standard one proposed by 
\cite{klinglesmith1970}. Thus, in order to  efficiently sample coefficients from the latter limb-darkening law one needs to derive an efficient  informative sampling scheme for it.

In this work, we aim to test how well these non-standard laws perform at retrieving transit parameters from transit lightcurves. We follow \cite{kipping2013} in order to 
derive an efficient sampling scheme for the standard form of the logarithmic and exponential laws, and show that the latter is actually \textit{always} non-physical 
in the sense proposed by \cite{kipping2013}. Then, we simulate transit lightcurves and perform simulations studies which extend the one done in EJ15 in several ways. 
First, we study not only the accuracy problem but also the precision issue on the retrieved parameters. Second, we study the non-standard limb-darkening laws by 
using a sampling scheme for the standard form of the logarithmic law derived in this work, the sampling schemes detailed in \cite{kipping2013} for the other two-parameter laws 
and \cite{kipping2015} for the three-parameter law. In addition, we also study the performance of the linear law, which has been used in recent years by several authors 
to parametrize the limb-darkening effect. We use our simulations to provide guidance as to when a particular law should be preferred, and provide code which, given the 
precision of a transit lightcurve and a set of transit and stellar parameters, helps to decide which law is the optimal to use.

This work is organised as follows. In \S2, we revisit the logarithmic and exponential laws in order to derive an efficient informative sampling strategy following the methods 
in \cite{kipping2013} for the standard forms of these laws. In \S3 we use those results in addition to the ones published by \cite{kipping2013} and \cite{kipping2015} in order 
to simulate transit lightcurves to test how well these laws perform in retrieving transit parameters both in terms of accuracy and precision. In \S4 we present a discussion and 
conclusions of our work.

% #################################################################################################################################
% #################################################################################################################################
% #################################################################################################################################
% ############################################# SECTION 2 #########################################################################
% #################################################################################################################################
% #################################################################################################################################
% #################################################################################################################################

\section{Efficient sampling of coefficients from the logarithmic and exponential laws revisited}

An informative and efficient sampling of limb-darkening coefficients is desirable in order to fit them to transit lightcurves. Motivated by this, 
\cite{kipping2013} introduced 
a procedure which first imposes physically motivated constrains on the center to limb stellar intensity profiles $I(\mu)$, namely, an everywhere 
positive intensity, $I(\mu)>0$,  and a decreasing intensity profile from center to limb, $\partial I(\mu)/\partial \mu>0$, 
which in turn impose constraints on the parameters of the law. With these constraints in hand, 
\cite{kipping2013} devised a triangular sampling strategy which only requires sampling two uniformly distributed numbers between 
0 and 1, $q_1$ and $q_2$, which through a transformation can be transformed to sample coefficients. This was done in that work for the most popular two-parameter limb-darkening laws with the exceptions of the 
exponential limb-darkening law, whose derived constraints on the coefficients were stated to be not enough 
to use the triangular sampling technique, and the typical form of the logarithmic law, which is 
different to the one used in \cite{kipping2013}. 

We derive here informative and efficient sampling strategy for the more typical form of the logarithmic law and then 
we study the exponential law, for which we conclude that deriving an efficient sampling strategy is actually 
impossible. We find that the law is fundamentally non-physical and, therefore, should not be used. 

\subsection{Efficient sampling from the logarithmic law}

As described in \cite{klinglesmith1970}, the logarithmic law is given by
\begin{eqnarray}
\label{loglaw} I(\mu) = 1 - l_1(1-\mu)-l_2\mu \ln\mu.
\end{eqnarray}
However, \cite{kipping2013} derived the constrains for a law similar (but 
not the same) to the original logarithmic law, given by
\begin{eqnarray*}
I_K(\mu) = 1 - A(1-\mu)-B\mu (1-\ln\mu).
\end{eqnarray*}

Given that this 
is not the law many authors have adopted we  derive here a sampling strategy for the original law given in eq. (\ref{loglaw}).  The constrain of an everywhere positive intensity profile implies
\begin{eqnarray*}
 l_1(1-\mu)+l_2\mu \ln\mu < 1  &\forall& 0<\mu<1.
\end{eqnarray*}
In order to ensure this condition, we need to find the extrema of the expression on the 
left-hand side. We note that a maximum of the expression is obtained only if $l_2<0$, 
while a minimum of the expression is obtained if $l_2>0$. In both cases, this happens 
at $\mu_e=\exp [(l_1-l_2)/l_2]$, where we have assumed $l_2 \neq 0$. If $l_2=0$, then 
we obtain the inequality $l_1(1-\mu)<1$, which implies $l_1<1$, a constraint we will see 
is valid for all the possible values of $l_2$. In the case $l_2>0$, because 
$\mu_e$ gives a minimum, 
$\mu\to 1$ and $\mu \to 0+$ maximise the expression in the desired range. The first limit 
gives the trivial constrain $1>0$, while the second limit gives the constrain $l_1<1$. In 
the case $l_2<0$, replacing $\mu_e$ in the expression leads to the constrain
\begin{eqnarray*}
\frac{l_1-1}{\mu_e} < l_2,
\end{eqnarray*}  
which, because $l_2<0$, again implies $l_1<1$. The condition 
for a decreasing intensity profile from center to limb, on the other hand, leads to the 
condition
\begin{eqnarray*}
l_1-l_2(1+\ln\mu) > 0  &\forall& 0<\mu<1.
\end{eqnarray*}
The left-hand side has different behaviour depending on the sign of $l_2$. If $l_2>0$, then 
the expression is convex and has no absolute minimum. It can be seen then that  the 
expression is always satisfied for $0<\mu < 1$. In the limit $\mu \to 1$, the expression leads 
to the condition $l_1>l_2$. For the $\mu \to 0^+$ limit no condition can be derived 
because the expression diverges. If $l_2<0$ the expression is concave, and 
it does not have an absolute maximum in $]0,1[$, the expression in the limit 
$\mu \to 1$ leads to $l_1>l_2$ just as in the first case, but in this case the expression is not valid 
for $\mu \to 0^+$, because it goes to $-\infty$. This implies that for $l_2<0$ the inequality will 
never be satisfied in the desired range and, thus, we need $l_2>0$ in order for the profile to be 
everywhere decreasing. In summary, the derived condition for an everywhere positive intensity profile is
\begin{eqnarray*}
l_1<1,
\end{eqnarray*}
while the conditions for an everywhere decreasing intensity profile from center to limb are
\begin{eqnarray*}
l_2>0,\\
l_1>l_2.
\end{eqnarray*}

\begin{figure}
\includegraphics{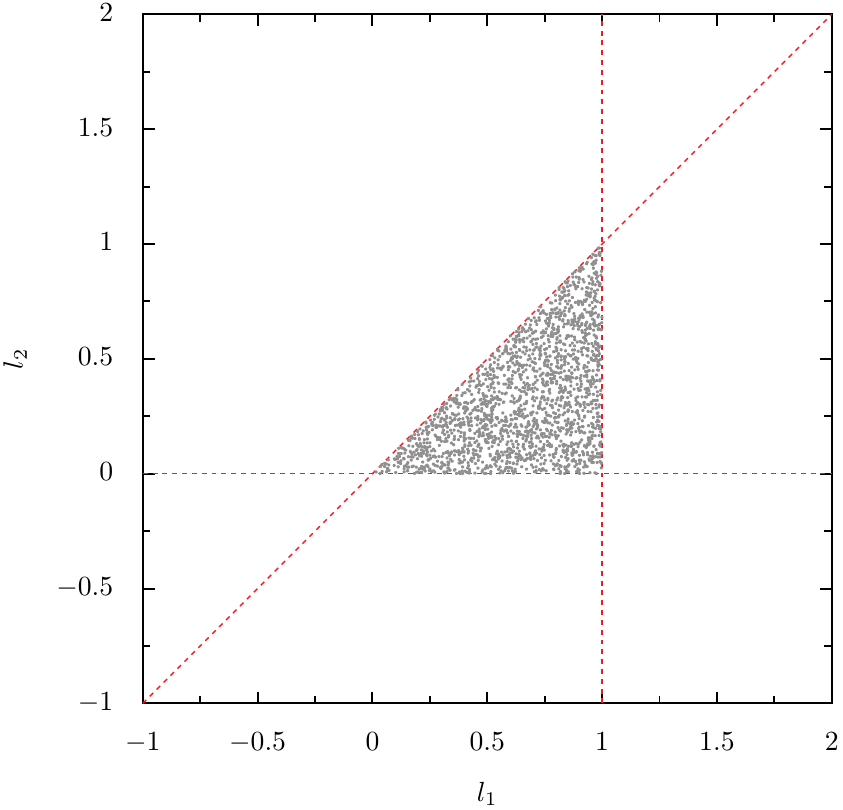}
\caption{Samples of the logarithmic limb-darkening coefficients that satisfy the derived constrains in 
this section out of $10^6$ uniformly sampled points between $-1<l_1<2$ and $-1<l_2<2$. Only 
$\sim 5\%$ of those samples satisfy such relations, shown here with red dashed lines.}
\label{triangular_sampling}
\end{figure}

Figure \ref{triangular_sampling} show these constrains geometrically. For illustration, $10^6$ points were uniformly sampled 
between $-1<l_1<2$ and $-1<l_2<2$ and a sample of those that satisfy these constrains were plotted. Only $5\%$ of them 
did satisfy such relations which demonstrates, as \cite{kipping2013} showed, the inefficiency of using such sampling strategy 
to draw physically plausible limb-darkening coefficients. This is the reason why we need to use the triangular sampling technique 
described in \cite{kipping2013}.

In order to use the triangular sampling technique, one needs to re-parametrize 
the constrains in order for them to be sampled on a right-angled triangle with this angle posed at the origin. A 
parametrisation that does this is the one with
\begin{eqnarray}
\label{v1}&v_1& = 1-l_1,\\
\label{v2}&v_2& = l_2.
\end{eqnarray}
If we now consider the transformations \citep[see][]{kipping2013}
\begin{eqnarray}
\label{v1q} &v_1& = \sqrt{q_1}q_2,\\
\label{v2q} &v_2& = 1 - \sqrt{q_1},
\end{eqnarray}
sampling $q_1$ and $q_2$ from uniform distributions between $(0,1)$ leads to a sampling with the desired constrains. Replacing the expressions in eqs. (\ref{v1q}) and (\ref{v2q}) in eqs. (\ref{v1}) and (\ref{v2}) gives 
\begin{eqnarray*}
l_1 &=& 1-\sqrt{q_1}q_2\\
l_2 &=& 1-\sqrt{q_1}
\end{eqnarray*}
which leads to the inverse equations
\begin{eqnarray*}
q_1 &=& (1-l_2)^2,\\
q_2 &=& \frac{1-l_1}{1-l_2}.
\end{eqnarray*}
We note that, as expected by construction, these relations return limb-darkening coefficients that are physically plausible for this law. Furthermore, we note these relations are different to the ones derived 
in \cite{kipping2013} as those are only applicable 
for his adopted form of the logarithmic law.

\subsection{The exponential law is non-physical}

We now study the exponential limb-darkening law. This law was 
introduced by \cite{claret2003}, and is given by
\begin{eqnarray*}
I(\mu) &=& 1-e_1(1-\mu)-e_2/(1-e^\mu).
\end{eqnarray*}
\cite{kipping2013} tried to apply the same methods used for the other two-parameter laws but observed that the triangular 
sampling technique was not applicable in this case because the physical constraints imposed were not 
able to yield a sufficient number of relations between the coefficients. However, apparently overlooked by \cite{kipping2013} is the fact that the exponential law will never
yield physically plausible coefficients for $0<\mu<1$, because the conditions of an everywhere positive and decreasing 
intensity profile from center to limb cannot be satisfied at the same time, as we now show.

We start by imposing an everywhere positive intensity profile for this law, which leads to the relation
\begin{eqnarray*}
e_1(1-\mu)+e_2/(1-e^\mu)<1 &\forall& 0<\mu<1,
\end{eqnarray*}
and the objective is to study the left-hand side. We note that this expression has a minimum if 
$e_2<0$, while it has a maximum if $e_2>0$. We see that in the first case, 
as $\mu \to 0^+$, 
the expression tends to $+\infty$, and therefore the expression cannot be satisfied for all $\mu$; this implies that 
$e_2$ cannot be less than zero. On the 
other hand, if we impose an everywhere decreasing intensity profile for this law from center to 
limb, this leads to the constraint
\begin{eqnarray*}
e_1 - e_2 \frac{e^\mu}{(1-e^\mu)^2}>0 &\forall& 0<\mu<1.
\end{eqnarray*}
The form of the left-hand side expression again depends on the value of $e_2$ in the same way 
as before, however, in this case there is no absolute maximum or minimum in $]0,1[$] and, thus, it suffices 
to study the expression at the borders. We note that in order for the expression to be satisfied 
for all the values of $\mu$, $e_2<0$, because only in this case the expression goes to $\infty$ 
when $\mu \to 0^+$. Because in order for the exponential law to have an everywhere positive intensity 
profile this condition was ruled out, this implies that both conditions cannot be met at the 
same time and, thus, there is no combination of coefficients $(e_1,e_2)$ that satisfy the physically 
plausible conditions suggested by \cite{kipping2013} for this law. This implies that the exponential 
law introduced by \cite{claret2003} is non-physical.

\begin{figure}
\includegraphics{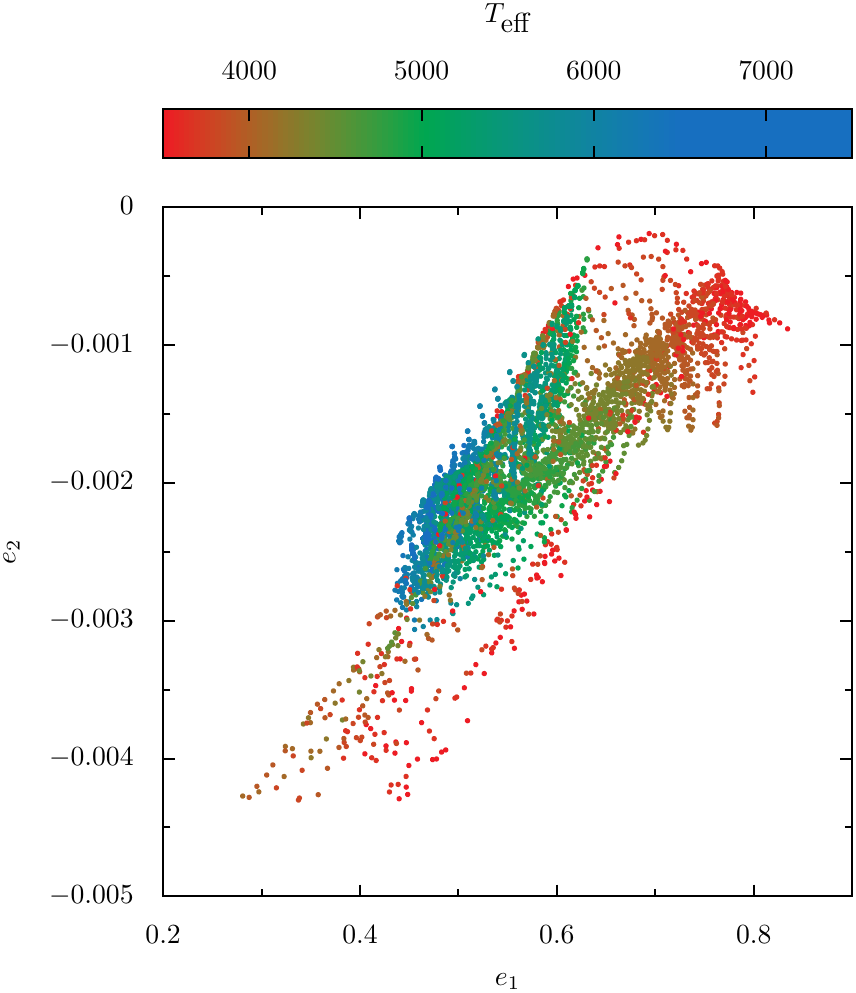}
\caption{Limb-darkening coefficients for the exponential law obtained using the methods described in 
EJ15 with the Kepler bandpass for all the stars in the ATLAS models with $T_\textnormal{eff}<9000$ K. As 
can be seen, $e_2<0$, which implies that all the fitted intensity profiles are not everywhere positive and, 
thus, non-physical.}
\label{exponential_ldcs}
\end{figure}

We note that typical values obtained by fitting the exponential law to model atmospheres give $e_2<0$, which means that 
most of the profiles fitted to model atmospheres do not have an everywhere positive intensity profile. To illustrate 
this, in Figure~\ref{exponential_ldcs} we plot all the coefficients fitted using the exponential law following the methods 
of EJ15 using the {\em Kepler} bandpass for model atmospheres with $T_\textnormal{eff}<9000$ K from  
the ATLAS models. In order to 
see at which value of $\mu$ the profile starts to be negative, we note that this happens as $\mu \to 0^+$. Here 
$1/(1-e^\mu) \approx -1/\mu$, which gives
\begin{eqnarray*}
I(\mu) &\approx& 1-e_1(1-\mu)+e_2/\mu.
\end{eqnarray*}
The intensity profile then, touches $I(\mu)=0$ around
\begin{eqnarray*}
\mu_0 \approx \frac{e_1-1 \pm \sqrt{(1-e_1)^2 - 4e_1e_2}}{2e_1}
\end{eqnarray*}
As shown Figure~\ref{exponential_ldcs}, $e_1\sim 0.6$, while $e_2 \sim - 2\times 10^{-3}$, which gives $\mu_0 \sim 0.005$, 
so the law starts giving negative values very close to (but not at) the limb, as we predicted. We note that typical transit observations are capable 
of sampling these points with high cadence observations as the ones done by the {\em Kepler} mission and the ones that will 
be done with the future TESS mission \citep{ricker2014}, which will have an almost certain chance of sampling points with $\mu\lesssim \mu_0$. 
Furthermore, as the generation of transit lightcurves for laws like the exponential requiere either numerical integration or the usage 
of  approximations that usually sample more points close to the limb \citep[which is the case for e.g., the \texttt{batman} code,][]{batman2015}, this 
law is doomed to produce numerical errors in transit lightcurves due to this fact. Because of these reasons, we advise not to use this law.

% #################################################################################################################################
% #################################################################################################################################
% #################################################################################################################################
% ############################################# SECTION 3 #########################################################################
% #################################################################################################################################
% #################################################################################################################################
% #################################################################################################################################

\section{Optimal selection of limb-darkening laws}

Having dealt with the sampling of limb-darkening coefficients from the 
logarithmic and exponential laws, we are now ready to turn to the study of the performance 
of the most popular limb-darkening laws (i.e., the linear, quadratic, logarithmic, square-root and 
three-parameter laws) both in terms of bias and precision in the 
retrieval of transit parameters. In particular, we focus on the performance of the retrieval of the planet-to-star 
radius ratio, $p=R_p/R_*$, the scaled semi-major axis, $a_R = a/R_*$ and the inclination, $i$, from transit lightcurves. 
In this section, we first study the biases introduced by using different limb-darkening laws, and then we study the 
bias-variance trade-off between different laws.

\subsection{Biases in the retrieval of transit parameters}

In order to study the biases introduced by different parametrizations of the limb-darkening effect, we perform a similar study 
as the one done in EJ15, but now (1) using the \texttt{batman} transit code described in \cite{batman2015}, (2) using the 
linear law, three different two-parameter laws (the quadratic, logarithmic and square-root laws) and the three-parameter 
law and (3) using a smaller number of steps in the planet-to-star radius ratio, in order to illustrate the evolution of the biases 
with this parameter in a cleaner way: we choose to study small ratios ($p=0.01$), medium ratios ($p=0.07$) and large ones 
($p=0.13$). 

As in EJ15, we simulate transit lightcurves for planets orbiting host stars with solar metallicity, 
gravity and microturbulent velocity, and $T_{\rm eff}$ between 3500 K and 9000 K, with the mentioned planet-to-star 
radius ratios, values of the scaled semi-major axis of $a_R = \{3.27, 3.92, 4.87, 6.45, 9.52, 18.18, 200\}$ and different 
impact parameters. The transits are simulated using a non-linear law in order to emulate ``real" intensity profiles, 
with the coefficients generated using the ATLAS models and the same methods as the ones used in EJ15 with the 
{\em Kepler} bandpass. As in this previous study, $1000$ in-transit points and $400$ out-of-transit points were simulated per 
lightcurve, with the initial times perturbed by a small amount in order for the results to be independent of the exact points at 
which the lightcurve is sampled (which anyways our simulations show has a negligible effect given the large amount of sampled 
points). The transits were then fitted using free limb-darkening coefficients with the different laws mentioned above.

For the simulations, the applied strategies to sample physically plausible coefficients from two-parameter laws 
are as follows, all of which use numbers $q_1 \in (0,1)$ and $q_2 \in (0,1)$:
\begin{itemize}
\item For the quadratic limb-darkening law, we fit for the numbers $q_1 = (u_1 + u_2)^2$ and $q_2 = u_1/2(u_1+u_2)$. 
To obtain the coefficients from those fitted numbers, we use the relations $u_1 = 2\sqrt{q_1}q_2$ and $u_2 =\sqrt{q_1}(1-2q_2)$, 
which are derived in \cite{kipping2013}.
\item For the square-root law, we fit for the numbers $q_1 = (s_1 + s_2)^2$ and $q_2 = s_2/2(s_1+s_2)$. 
To obtain the coefficients from those fitted numbers, we use the relations $s_1 = \sqrt{q_1}(1-2q_2)$ and $s_2 =2\sqrt{q_1}q_2$, 
which are derived in \cite{kipping2013}.
\item For the logarithmic law, we fit for the numbers $q_1 = (1-l_2)^2$ and $q_2 = (1-l_1)/(1-l_2)$. 
To obtain the coefficients from those fitted numbers, we use the relations $l_1 = 1-\sqrt{q_1}q_2$ and $l_2=1-\sqrt{q_1}$, 
which are derived in \S 2.1.
\end{itemize}
For the three-parameter law, we use the formalism and code described in \cite{kipping2015}, while for the linear law we simply 
sample coefficients between $0$ and $1$, which give intensity profiles which are both always positive and which show a strictly 
decreasing intensity profiles from center to limb.  The code used to generate our simulations is available in GitHub\footnote{\url{http://www.github.com/nespinoza/ld-exosim/}}, 
and can be used to perform analyses in a finer grid than the ones published here. 

\subsubsection{The case of central transits}

\begin{figure*}
\includegraphics[scale=0.67]{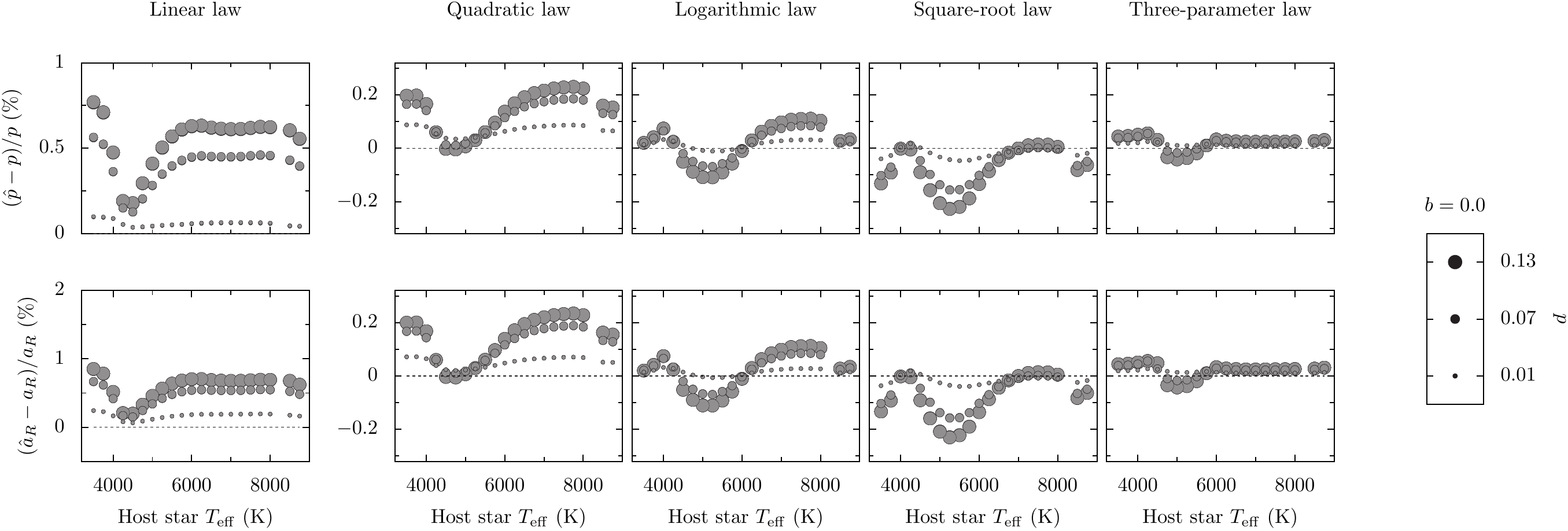}
\caption{Results of fitting transit lightcurves as described in \S 3.1 with the limb-darkening coefficients as free parameters. Note that the scale shown for the linear law is different 
than the one used in the figures for the other laws. The size of the points is proportional to the the input values of $p$.}
\label{central_sims}
\end{figure*}

The results of the simulations for central transits (i.e., $b=0$) are shown in Figure~\ref{central_sims}, with the sizes of the points proportional to the 
input values of $p$. The upper panels show the biases introduced on the planet-to-star radius ratio, $p$, and the lower panels show the biases introduced 
on the scaled semi-major axes, $a_R$. From left to right, the panels show the biases introduced by using a linear law, a quadratic law, a logarithmic law, a 
square-root law and a three-parameter law. The dependence with the input values of $a_R$ are not shown as the bias seems to be almost independent of 
this parameter in this case\footnote{This can be also seen by looking at the lower panels of Figure~10 in EJ15, which give the same results as the ones shown here for 
the quadratic law. We note that this implies good agreement between the simulations done with our implementation of the \cite{ma2002} codes for the transit light 
curve generation and the \texttt{batman} code}. 

As can be seen from the figure, different limb-darkening laws have different behaviours in retrieving $p$ and $a_R$, with a dependence on the input value of 
$p$, and also depending on the temperature of the host star. First, it is interesting to note that the linear law is by far the worst of the laws in terms of bias. As 
expected from its very simple form, the biases on the retrieved transit parameters are larger than the ones observed on the two and three-parameter laws for 
almost all the stellar temperatures except for $T_\textnormal{eff}\sim 4500$, where the profile seems to be a decent  approximation to the real intensity profile (yet poorer than the ones provided by the 
other laws). A second feature to note on our results is that the logarithmic law does a very good job at retrieving the 
parameters of small planet-to-star radius ratios, even at the level of the three-parameter law for stars with $T_\textnormal{eff}<6000$ K, and also for medium 
and large planet-to-star radius ratios with host-stars with $T_\textnormal{eff}<4000$ K. This makes this law a very good choice for present and future transit characterisation around M-dwarfs. The quadratic law, on the other hand, is only the best for host-stars between 
$4250<T_\textnormal{eff}<5500$ K, with the logarithmic law being the best in the $5500<T_\textnormal{eff}<6250$ K range, and the square-root law 
being the best for planets around host stars with temperatures between $6250<T_\textnormal{eff}<8000$ K. Hotter than this, the best option in terms 
of accuracy is the three-parameter law.

It is interesting to emphasise that although the biases introduced by the quadratic law might seem to be small, as mentioned in the introduction and in 
EJ15 they are significant for several candidate and confirmed exoplanets. Choosing the right law can help to achieve the same 
level of accuracy than the level of precision that {\em Kepler} achieves for most planets. For example, accuracies on the order of $\sim 0.01\%$ can 
be achieved for $p$ with a careful selection of a two-parameter law, and exoplanets with precisions better than that according to a query to the Nasa Exoplanet 
Archive\footnote{Query done on 29/09/2015.} form only $0.58\%$ of all candidate exoplanets and $1.2\%$ of the confirmed exoplanets. For the remaining 
exoplanets requiring a better accuracy, a new law might have to be created, as this is also the limit of the three-parameter law (which has an accuracy 
on $p$ around $\sim 0.01\%$ for stars with temperatures larger than $T_\textnormal{eff}=6000$ K). Studying new limb-darkening laws, however, 
is out of the scope of the present work.

\subsubsection{The case of non-central transits}

\begin{figure*}
\includegraphics[scale=0.67]{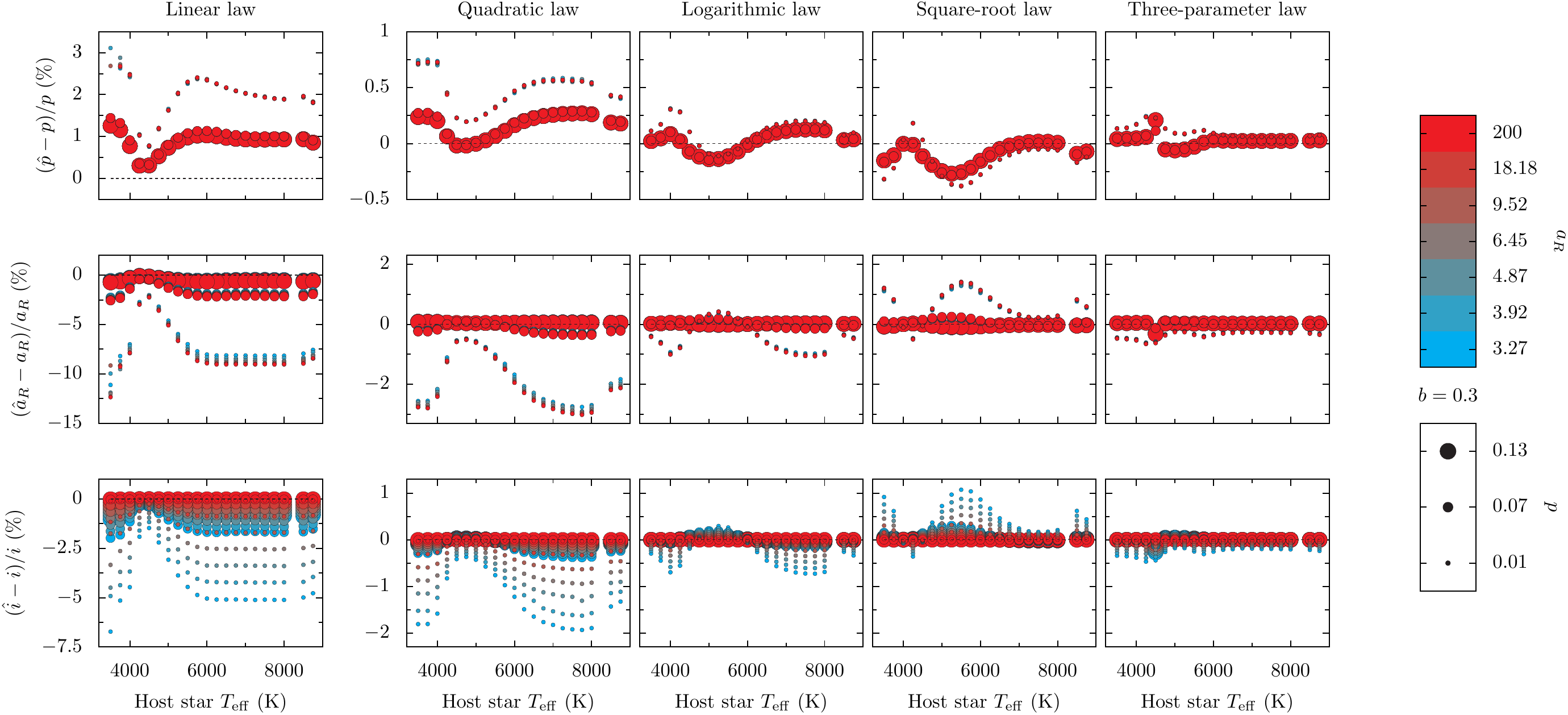}
\caption{Results of fitting transit lightcurves as described in \S 3.1 with the limb-darkening coefficients as free parameters, but for low impact 
parameter transits ($b=0.3$). Note that the scale shown for the linear law is different than the one used in the figures for the other laws. The 
size of the points represent the input values of $p$, while their color represent the input values of $a_R$.}
\label{small_b_sims}
\end{figure*}

\begin{figure*}
\includegraphics[scale=0.67]{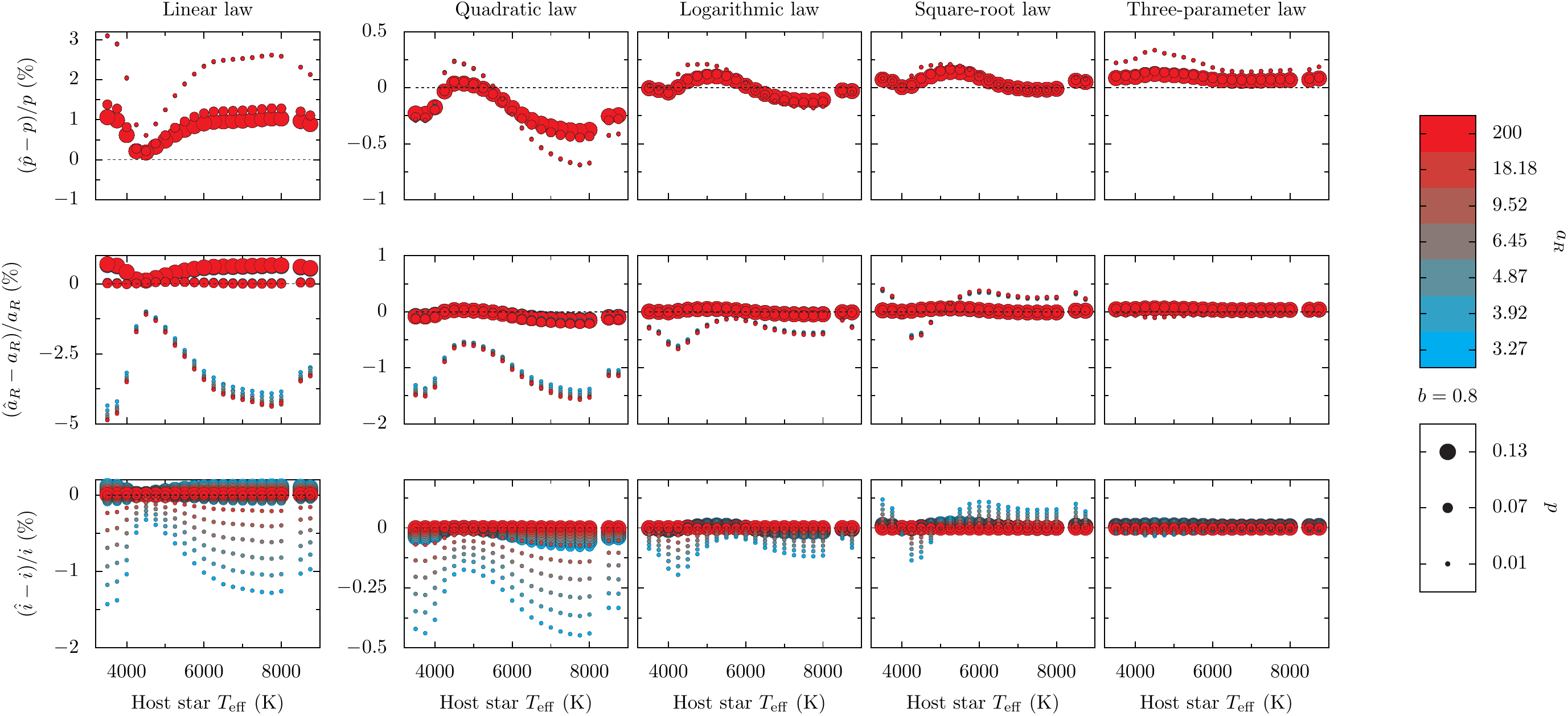}
\caption{Results of fitting transit lightcurves as described in \S 3.1 with the limb-darkening coefficients as free parameters, but for high impact 
parameter transits ($b=0.8$). Note that the scale shown for the linear law is different than the one used in the figures for the other laws. The size 
of the points represent the input values of $p$, while their color represent the input values of $a_R$.}
\label{big_b_sims}
\end{figure*}

Figures~\ref{small_b_sims} and \ref{big_b_sims} show the results for low ($b=0.3$) and high ($b=0.8$) input impact parameters respectively. In these cases, 
we see that the biases are in general worse for smaller values of $p$ and $a_R$, i.e., for a given stellar radius, for smaller, close-in planets. We can 
also see that, again, the worst law to choose in terms of accuracy is the linear, which can give rise to biases on the order of $\sim 3\%$ for small planet-to-star 
radius ratios. However, it is evident from the results that a careful selection of the law to be used can allow one to minimize the biases on the retrieved transit 
parameters. For example, if the objective were to obtain $p$ with minimum bias for a low impact-parameter transit, then for host stars colder than $4000$ K 
the best option is to use either the logarithmic or the three-parameter law, which show biases on the order of $\sim 0.1\%$. In contrast, using the quadratic law 
for transits around those host stars can give rise to biases on the order of $\sim 0.75\%$, i.e., almost an order of magnitude larger. 

Overall, as expected, the law that retrieves the parameters with minimum bias in low impact-parameter transits is the three-parameter law. However, for 
high impact-parameter transits of small planets, the logarithmic and square-root law seem to outperform the three parameter law in general if the objective is to retrieve 
a minimum bias estimator for $p$. This is due to the fact that the three-parameter law is so flexible that it allows for small distortions of the transit lightcurve that, 
in this case, can mimic variations that were actually attributable to $p$, adding a small but important bias on the retrieved parameter. However, for the other 
parameters ($a_R$ and $i$) the three parameter law seems to be the best choice in terms of accuracy.

\subsection{The bias-variance trade-off}

\begin{figure}
\includegraphics[scale=0.73]{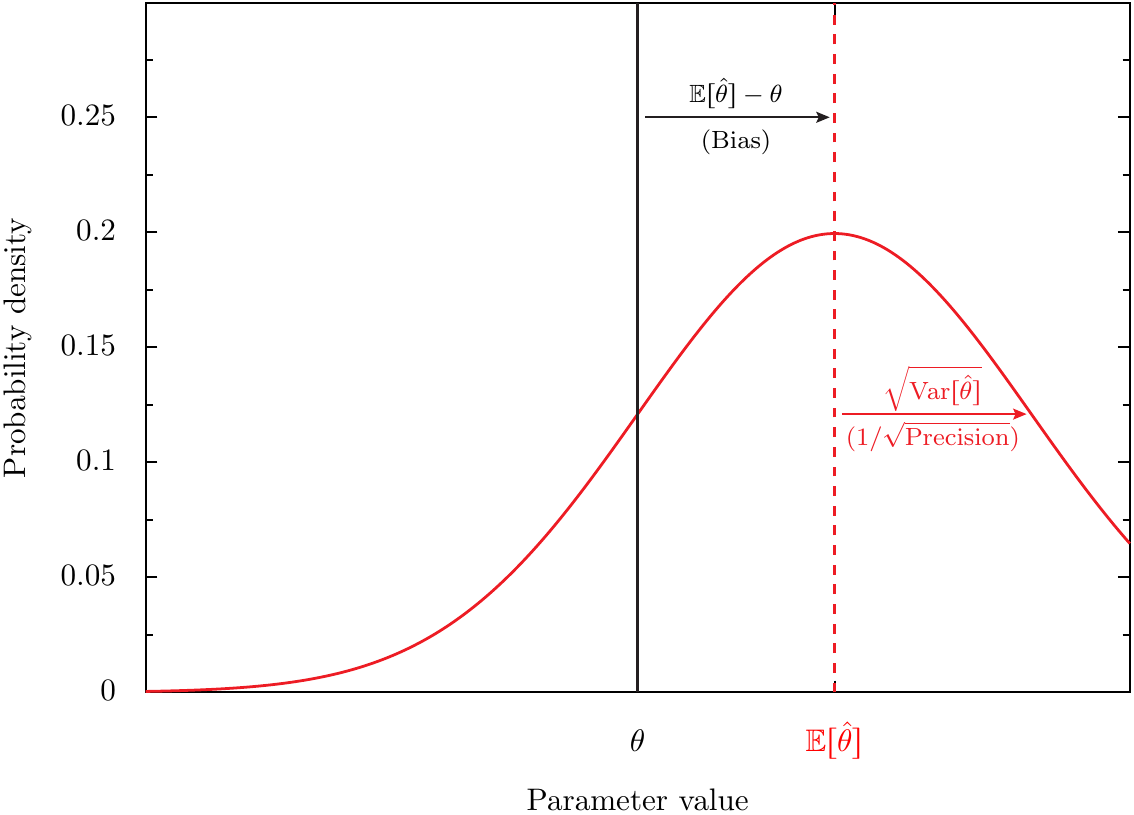}
\caption{Sketch of each of the terms that define the bias and the precision of a given estimator. The probability density of the estimator $\hat{\theta}$ of 
the parameter $\theta$ is depicted in red, while the real, underlying value of the parameter $\theta$ is depicted with a black solid line. Ideally, one would 
want a low bias, high precision estimator.}
\label{bias-variance-plot}
\end{figure}

From the results on the past sub-section, one can see that for a given set of stellar and transit parameters, a law can be chosen in order to minimize the 
biases in the retrieved transit parameters from the transit light curve. However, an interesting question is whether the precision of a given lightcurve can 
actually provide enough precision in order to use a more complex parametrization of the limb-darkening effect and, depending on the answer, what is 
the cost. Fitting a complex law to a noisy light curve will definitely lower the precision of the retrieved transit parameters due to the high flexibility of the 
model, while fitting a law that is too simple for a very precise lightcurve will most definitely lead to larger 
biases, due to the strictness of the chosen model. This problem is known as the bias-variance trade-off. Probabilistically, for a given estimator $\hat{\theta}$ of 
the parameter $\theta$, we will be interested in its variance (whose inverse defines the precision), $\textnormal{Var}[\hat{\theta}]$, and its bias, defined as 
$\mathbb{E}[\hat{\theta}]-\theta$, where $\mathbb{E}[\cdot]$ defines the expectation value and $\textnormal{Var}[\cdot]$ is the variance operator. The bias, on 
one hand, defines how far on average is the estimator from the parameters' true value, while the variance defines the spread around the expected value of the 
estimator (see Figure \ref{bias-variance-plot}). One ideally wants a low-bias and high-precision estimator, which is both close to the real value of the parameter on average and, 
at the same time, has a low spread around such value. The problem is that in real applications one has the possibility to choose from various estimators with different 
ammounts of bias and variance, and thus choosing among them is, at first, not trivial. This is when the mean-squared error metric comes in handy. The mean-squared 
error (MSE) of an estimator $\hat{\theta}$ of the parameter $\theta$ is defined as
\begin{eqnarray*}
\textnormal{MSE} &=& \mathbb{E}[(\theta - \hat{\theta})^2].
\end{eqnarray*}
From the definition of the MSE, it can be seen that it is a measure of the square of the distance from the estimator to the parameters' ``true" value. This provides a metric among a given suite of estimators for a given parameter: the one that gives the minimum mean-squared error 
provides, on average, the one that is closer to the underlying ``true" value on a mean-square sense. Furthermore, it is easy to show that the MSE can be written 
as
\begin{eqnarray*}
\textnormal{MSE} &=& \mathbb{E}[(\theta - \hat{\theta})^2], \\
&=& (\mathbb{E}[\hat{\theta}]-\theta)^2 + \textnormal{Var}[\hat{\theta}],\\
&=& \textnormal{Bias}^2 + \textnormal{Variance}.
\end{eqnarray*}
This implies that given the bias and the variance of an estimator, the MSE can be easily calculated. In our specific application, in the past sub-section we have already 
shown and studied the biases that each limb-darkening law (i.e., different estimators) imply on the retrieved transit parameters through simulation. We have thus already
calculated the floor of the MSE for each law, and it only remains to study the variances implied by those limb-darkening laws, which is a function of the noise of the lightcurve 
used to retrieve the transit parameters.

\begin{figure*}
\includegraphics[scale=0.9]{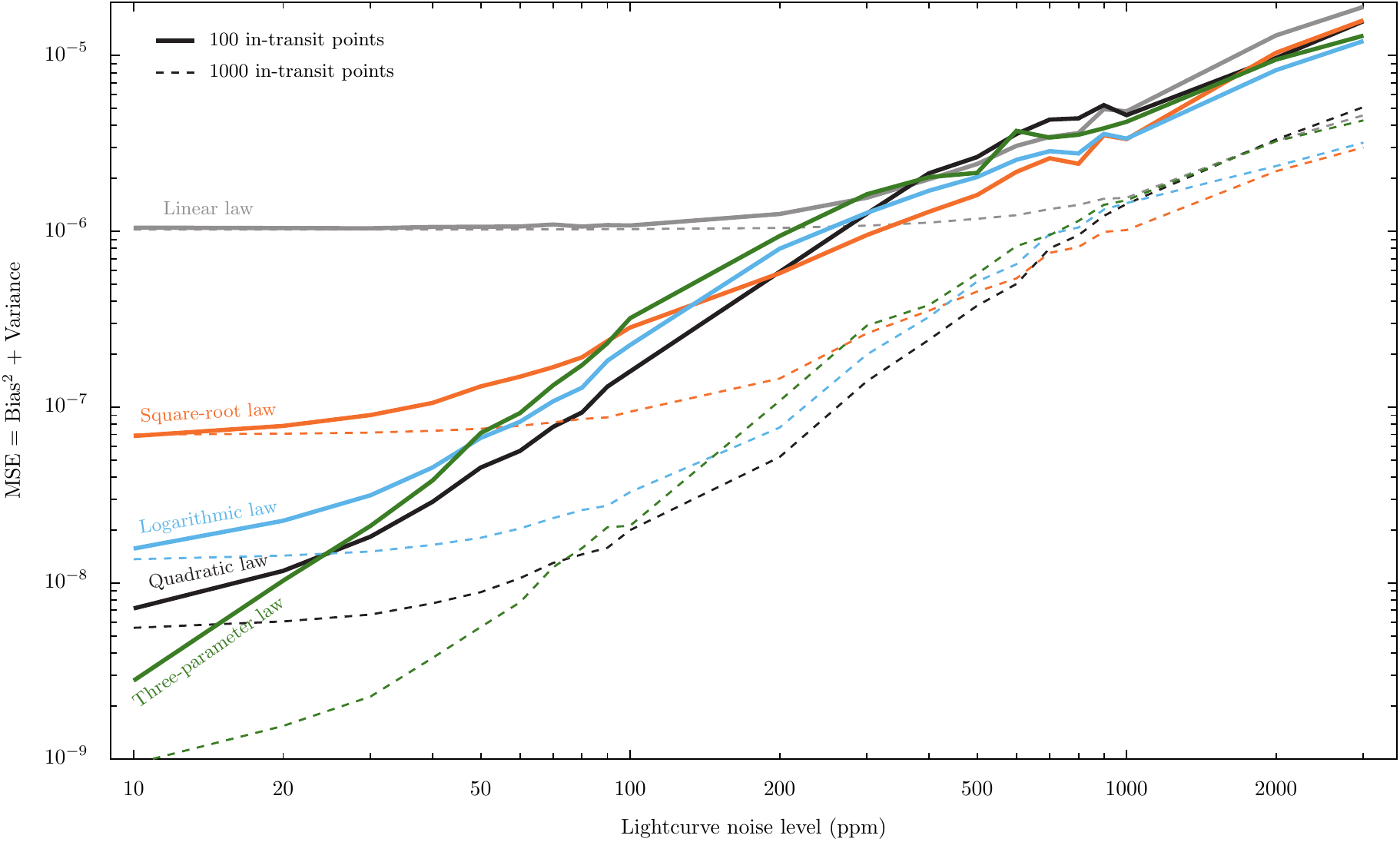}
\caption{Results of our lightcurve fits to noisy lightcurves of a typical Hot-Jupiter in order to study the bias-variance trade-off of different limb-darkening laws on the planet-to-star radius 
ratio for lightcurves with $100$ and $1000$ in-transit points. The MSE for different laws is shown as a function of the lightcurve noise level.}
\label{biasvariance}
\end{figure*}

Given that the study of the mean-squared error is complex from a mathematical point of view for our case, we study it through simulations similar to the ones done in the past sub-section, 
but this time also adding different amounts of noise to the lightcurves in order to derive the variances of the estimators for a given level of noise. We take a Monte Carlo approach in order to study 
the MSE. As a case study, we take that of a typical Hot-Jupiter ($p=0.1$, $a_R = 10$) around a Solar-type star ($T_\textnormal{eff}= 5500$ K) on a low impact-parameter orbit ($b=0.3$), 
and we focus specifically on the retrieval of the planet-to-star radius ratio, which is one of the most interesting parameters retrieved from transit lightcurves. We simulate a suite of 
lightcurves with noise levels ranging from $10$ ppm to $3000$ ppm with different amount of in-transit points. We show our results for $N=100$ and $N=1000$ in-transit 
points, which span a range of possible scenarios for both space based (e.g., \textit{Kepler}) and ground-based observations. For each case, $300$ lightcurves are generated, 
and all of them are fitted in the same way as it was done for their noiseless counterparts in the past sub-section. From the fit to each of the $300$ lightcurves on each case, 
we take all the estimated planet-to-star radius ratios and estimate the variance of the sample. The results of our simulations are shown on Figure~\ref{biasvariance}, where we plot the 
MSE as a function of the noise of the simulated lightcurves for each of the limb-darkening laws used to retrieve the parameter. 

It is evident from our results how the MSE evolves from being variance-dominated for high-noise lightcurves (rightmost part of the plot) to bias-dominated for low-noise 
lightcurves (leftmost part of the plot). This evolution is most evident for the linear law: for $N=100$ in-transit points, for example, the linear law attains the bias floor around 
$\approx100$ ppm. From here and for more precise lightcurves, the MSE of this law attains the constant value $\sim 10^{-6}$, which corresponds to the bias floor of 
$10^{-3}$, which for $p=0.1$ corresponds to the $1\%$ bias observed on our simulations on the past sub-section (see Figure~\ref{small_b_sims}). 
Note that from this plot, the optimal law in terms of the MSE at each noise level can be selected with ease. For example, in our simulation, for $N=100$ in-transit points, 
the logarithmic law is the best for the retrieval of the planet-to-star radius ratio in a MSE sense for lightcurves with noise levels between $\approx3000$ ppm to $\approx 1000$ ppm. 
At this last noise level, we see that the square-root law crosses the line of the logarithmic law, and this latter law attains the lowest MSE until $\approx 200$ ppm, where the 
quadratic law starts to get the lowest MSE. The quadratic law is thus the best law to use from $\approx 200$ ppm until lightcurves with precisions of $\approx25$ ppm, a noise level at 
which the three-parameter law takes over and attains the lowest MSE. Note that the number of in-transit points changes the  points at which the transition among different limb-darkening 
laws occur due to the fact that the bias floor is attained faster as a consequence of the increased number of points. For example, the transition between the quadratic and three-parameter 
laws occurs for lower precision lightcurves for the case of $1000$ in-transit points (at $\approx 75$ ppm).

From our case study, we propose that the selection of the optimal limb-darkening law in a MSE sense can be estimated by simple simulation:  for a given observational setup in terms of cadence and noise, simply calculate the MSE obtained 
by all the laws and select the one that gives the minimum MSE. However, we note that different transit geometries will give rise to different optimal limb-darkening laws at a given 
lightcurve precision and, thus, these simulations must be performed on a case-by-case basis. Of course, for this case-by-case study a set of transit parameters are needed as 
inputs, which is a bit circular as this is what we are trying to extract from a transit lightcurve. However, as we learned from our results on this and the past sub-sections, it is important 
to remember that the functional relation of the transit parameters and the biases/variances is smooth and slow. Therefore, one can first fit with one's favourite law, obtain a set of 
transit parameters, and use those as input in order to calculate the MSE of different limb-darkening laws and then retrieve the final transit parameters using the law that gives 
the lowest MSE. We provide code at Github with which this process can be easily done for a given set of transit and stellar parameters, along with a given lightcurve noise level and 
number of in-transit points. 

\section{Discussion \& Conclusions}

In this work we have explored the different biases and precisions on retrieved transit parameters incurred by the usage of different limb-darkening laws. In particular, we 
explored the biases introduced by using the linear law, three two-parameter laws (the quadratic, logarithmic and square-root laws) and the three-parameter law of \citet{sing2009}, 
and defined a way to select the optimal law for a given lightcurve precision and number of in-transit points. We showed that a careful selection of the limb-darkening law to be 
used in a given dataset can allow one to retrieve optimal estimators for the transit parameters from transit lightcurves. In general, as shown in \S3, two-parameter laws have ranges 
in which they perform at the same level or even better than the three-parameter law and, because of this, we encourage their use. We also discourage the usage of the exponential 
law which, as shown in \S2.2 is unphysical, and can lead to numerical errors if used on real data where sampling of points close to the limb is not uncommon.

The question of when to use each of those laws is an important one. In particular, we have explored the relation between the ``real", underlying 
transit parameters, the stellar host properties and the bias/variance on the retrieved transit parameters. We have stated no specific rule regarding when to use each 
law, but rather provided guidelines based on our simulations. This is a choice made on purpose, as the selection of which law to use has to be done in a 
case-by-case basis, a task for which we provide code with which one can obtain the results of the same simulations performed on our work for specific 
cases \footnote{\url{http://www.github.com/nespinoza/ld-exosim/}}. It is important to note that in this work we have used the 
{\em Kepler} bandpass to study the relation between the biases/variances on the retrieved transit parameters and the stellar host properties, but different bandpasses should give 
rise to  different functional relations because they give rise to different model intensity profiles. Limb-darkening coefficients have to be calculated for each  particular case, which is straightforward to do using the algorithms described in EJ15\footnote{\url{http://www.github.com/nespinoza/limb-darkening/}}.

\section{Acknowledgmenets}

We thank the referee for insightful comments that greatly improved 
this work. N.E. is supported by CONICYT-PCHA/Doctorado Nacional. A.J. acknowledges 
support from FONDECYT project 1130857, the Ministry
for the Economy, Development, and Tourism Programa Iniciativa
Cient\'ifica Milenio through grant IC 120009, awarded to the
Millennium Institute of Astrophysics (MAS), and from BASAL CATA
PFB-06.

\bsp

\label{lastpage}

\end{document}